\documentclass[11pt,a4paper]{article}
\usepackage{amssymb}
\usepackage{graphicx}
\usepackage{amsmath}
\usepackage{makeidx}
\usepackage{indentfirst}
\usepackage[T1]{fontenc}
\usepackage[utf8]{inputenc}
\usepackage{authblk}

\newcounter{resultnum}[section]
\newcounter{conclusionnum}[section]
\newcounter{conditionnum}[section]
\newcounter{conjecturenum}[section]
\newcounter{examplenum}[section]
\newcounter{exercisenum}[section]
\newcounter{lemmanum}[section]
\newcounter{notationnum}[section]
\newcounter{theoremnum}[section]
\newcounter{definitionnum}[section]
\newcounter{corollarynum}[section]
\newcounter{remarknum}[section]
\newcounter{propositionnum}[section]
\newcounter{acknowledgementnum}[section]
\newcounter{algorithmnum}[section]
\newcounter{axiomnum}[section]
\newcounter{casenum}[section]
\newcounter{claimnum}[section]
\newcounter{summarynum}[section]
\newcounter{problemnum}[section]
\begin{document}

\title{Covariant Renormalizable Modified and\\ Massive Gravity Theories on (Non) Commutative Tangent Lorentz Bundles}

\date{April 23, 2013}
\author{\textbf{Sergiu I. Vacaru} \thanks{sergiu.vacaru@uaic.ro;\newline http://www.uaic.ro/uaic/bin/view/Research/AdvancedTheoretical}}

\affil{\small Rector's Office, Alexandru Ioan Cuza University,  Alexandru Lapu\c sneanu \newline street, nr. 14,  UAIC -- Corpus R, office 323;  Ia\c si,\ Romania, 700057 }

\renewcommand\Authands{ and } %

\maketitle

\begin{abstract}
The fundamental field equations in modified gravity  (including general relativity; massive and bimetric theories; Ho\v rava-Lifshits, HL; Einstein--Finsler gravity extensions etc) posses an important decoupling property with respect to nonholonomic frames with 2 (or 3) +2+2+... spacetime decompositions. This allows us to construct exact solutions  with generic off--diagonal metrics depending on all spacetime coordinates via generating and integration functions containing (un-) broken symmetry parameters. Such nonholonomic configurations/ models have a nice ultraviolet behavior and seem to be ghost free and (super) renormalizable in a sense of covariant and/or massive modifications of HL gravity. The apparent noncommutativity and breaking of Lorentz invariance by quantum effects can be encoded into fibers of noncommutative tangent Lorentz bundles for corresponding "partner" anisotropically induced theories. We show how the constructions can be extended to include conjectured covariant reonormalizable models with massive graviton fields and effective Einstein fields with (non)commutative variables.

\vskip5pt

\end{abstract}

\newpage 




\section{Introduction}

As there is no hint how a quantum gravity theory (QG) can be formulated and
verified for high energies, it is important to explore and compare certain
promising models which can be unified into a unified commutative and
noncommutative geometric formalism related to a general method of
construction exact solutions. In this paper, we analyze  three intensively
developing approaches to quantum gravity. The aim is to show that we are
able to formulate any such a quantum theory as a consistent, predictive and
observational (in modern cosmology) one with phenomenological implications
in high energy particle physics. The first approach is based on a proposal
\cite{horava} (the so--called Ho\v rava -- Lifshitz, HL, gravity) to
consider Lorentz non--invariant theories with scaling properties of space, $%
x^{i},$ and time, $t,$ coordinates re--parameterized in the form $(\mathbf{x}%
,t)\rightarrow (b\mathbf{x},b^{z}t),$ where $z=2,3,...$ This modifies the
ultraviolet (UV) behavior of the graviton propagator by changing $1/|\mathbf{%
k}|^{2}\rightarrow 1/|\mathbf{k}|^{2z},$ where $\mathbf{k}$ is the spacial
momentum. Such constructions were originally performed with the lack of full
diffeomorphysm invariance and resulted in the impossibility to exclude
completely certain un--physical modes, see critical remarks in a series of
works \cite{moffat}. Latter,  a covariant renormalizable
gravity model developing a HL--like gravity to full diffeomorphysm
invariance  was constructed \cite{odints1}. The main idea was to broke the
Lorentz--invariance of the graviton propagator by introducing a
non--standard coupling with an unknown fluid. Such theories with possible
physical transverse modes seem to be (super--) renormalizable and certain
applications in modern accelerating cosmology where provided.

The second approach is related to a recent substantial progress made with
massive/ bimetric gravity theories \cite{rhamhr} (for historical remarks,
motivations and review of results and applications, see also \cite{odints}
and references therein). There are involved two metrics (for certain models,
it is considered also a second connection) when the second one describe an effective exotic
matter induced by massive gravitons. Such theories do not suffer from the
ghost instability in a well defined perturbation theory (away from the
decoupling limit). In a more general context (third approach), a QG must
include noncommutative type configurations, for instance, induced by Schr%
\"{o}dinger type uncertainty relations with the Planck quantum constant \cite%
{vnonc}. We can encode such theories into certain geometric models on (co)
tangent bundles to spacetime manifolds (in standard form, there are used
Lorentz manifolds and/or various modifications with nontrivial torsion,
extra dimensions etc). It is possible to consider commutative and
noncommutative modified gravity theories admitting exact off--diagonal
solutions \cite{vexsolbranes} with classical and quantum variables \cite%
{vquant} subjected to nonholonomic constraints and with possible
generalizations to Finsler--Lagrange--Hamilton variables \cite{vfinsl}.

In a series of works \cite{vexsolbranes,vnonc,vfinsl}, there were provided
detailed proofs and examples when the gravitational field eqs in general
relativity, GR, and various modifications (with commutative and
noncommutative variables, extra dimensions, massive terms, modified Lagrange
densities, anistoropic dependencies on velocity/momentum type variables,
string like and brane theories etc) can be decoupled with respect to certain
classes of nonholonomic frames. The solutions for such nonlinear systems of
(generalized/modified) Einstein equations can be
constructed in very general forms. Such spacetimes are characterized by the
generic off--diagonal metrics which can not be diagonalized via coordinate
transforms and depend on all spacetime coordinates via corresponding
generating and integration functions and possible (broken, or preserving)
symmetry parameters. Possible nontrivial torsion configurations can be
nonholonomically constrained to the Levi--Civita ones (with zero torsion).
Choosing necessary types of generating/ integration functions and parameters
and performing corresponding deformations of frame, metric and connection
structures, we can model effective nonlinear interactions, (modified)
massive gravity effects, with scaling properties and local anisotropies,
which can be renormalizable.

In this work, we study (non) commutative massive gravity theories which can
equivalently modelled as Lagrange density modified ones and encoded in (effective)
Einstein spaces and generalizations on tangent Lorentz bundles. We shall
state the conditions when (effective/modified) Einstein equations transform
nonlinearly, by imposing corresponding classes of nonholonomic constraints,
to nonlinear systems of partial differential eqs (PDE) with parametric
dependence of solutions which under quantization \textquotedblright
survive\textquotedblright\ and stabilize in some rescaled/anisotropic and
renormalized forms.

 The paper is organized as follows. In section \ref{sec2}, we state the
 actions for  modelling commutative modified and effective
theories of gravity on Lorentz manifolds. There are provided the
corresponding generalized gravitational field equations. We consider also
nonholonomic deformations of such models on (non) commutative tangent
bundles determined by Sch\"{o}dinger type complex/noncommutative relations.
Section \ref{sec3} is devoted to a geometric method for decoupling and
integrating the gravitational field equations with respect to nonholnomic
frames. Then we consider off--diagonal solutions mimicking such effective
theories for nonstandard perfect fluid coupling in section \ref{sec4}. We
speculate how modified (non) commutative massive gravity theories can be
renormalize in a covariant HL sense using effective Einstein and/or
Einstein--Finsler type spaces.  Finally, we conclude the paper in section \ref{sec5}.

\section{(Non) Commutative Modified Massive Gravity} \label{sec2}
\subsection{Actions for equivalent commutative gravity theories}
We consider four equivalent models determined by actions
{\small
\begin{equation}
S=\frac{1}{16\pi }\int \delta u^{4}\sqrt{|\mathbf{g}_{\alpha \beta }|}\
\mathcal{L},\mbox{ for }\mathcal{L=}\ ^{[i]}\mathcal{L},[i]=1,2,3,4;
\label{act}
\end{equation}%
}
(two $f$--modified, a massive  and an effective Einstein gravity theories), where
\begin{eqnarray*}
\ ^{[1]}\mathcal{L} &=&\widehat{f}(\widehat{R})-\frac{\mathring{\mu}^{2}}{4}%
\mathcal{U}(\mathbf{g}_{\mu \nu },\mathbf{K}_{\alpha \beta })+\ ^{m}L,~^{[2]}%
\mathcal{L}=~^{s}\widetilde{R}+\widetilde{L}, \\
\ ^{[3]}\mathcal{L} &=&R+L(T^{\mu \nu },R_{\mu \nu }),\ ^{[4]}\mathcal{L}=f(%
\breve{R})+\ ^{m}L.
\end{eqnarray*}%
Such theories are modelled on a four dimensional (4d) pseudo--Riemannian
manifold $\mathbf{V}$ with physical metric $\mathbf{g=\{g}_{\mu \nu }\}$ of
signature $(+,+,+,-).$ In GR, $\mathbf{V}$ is defined as a Lorentz manifold
with a corresponding axiomatic and physical interpretation which can be
extended to Einstein -- Finsler like models on tangent Lorentz bundles \cite%
{vfinsl}. Commutative modifications of gravity theories are considered for
the same metric structure but with different connections and/or Lagrange
densities. For massive gravity models with "small" graviton mass $\mathring{%
\mu}$, the physical spacetime $\mathbf{V}$ is modelled in a similar form but
with an additional bimetric structure related to a "sophisticate" potential of
graviton $\mathcal{U},$ see details in \cite{rhamhr}; it is involved a
fiducial metric determined by a tensor field $\mathbf{K}_{\alpha \beta }$,
see details in Refs. \cite{rhamhr}. The symbols $\ ^{m}L$ and $L$ are used
respectively for Lagrange densities of matter fields and effective
matter.

In above formulas, $\widehat{R}$ is the scalar curvature for an auxiliary
(canonical) connection $\widehat{\mathbf{D}}$ uniquely determined by $%
\mathbf{g}$ for a conventional nonholonomic horizontal, h, and vertical, v,
splitting following two conditions: 1) It is metric compatible, $\widehat{%
\mathbf{D}}\mathbf{g}=0,$ and 2) its $h$- and $v$-torsions are zero
(but there are nonzero $h-v$ components of torsion $\widehat{\mathcal{T}}$
completely determined by $\mathbf{g).}$ Geometrically such a decomposition
(splitting) is stated as a Whitney sum
\begin{equation}
\mathbf{N}:T\mathbf{V=}h\mathbf{V\oplus }v\mathbf{V,}  \label{whitney}
\end{equation}
see details in \cite{vexsolbranes}.\footnote{%
For a conventional $2+2$ splitting, the coordinates can be labelled in the
form $u^{\alpha }=(x^{i},y^{a}),$ or $u=(x,y),$ with indices $i,j,k...=1,2$
and $a,b,...=3,4.$ Boldface symbols are used in order to emphasize that
certain geometric/physical objects and/or formulas are written in a
N--adapted form. There will be considered left up/low indices as labels for
certain classes of geometric/physical objects. We shall use the Einstein
rule on summation of repeating right up-low indices if the contrary will be
not stated.} The "priority" of the connection $\widehat{\mathbf{D}}$ is that
it allows to decouple the field equations in various gravity theories which
allows us to construct exact solutions in very general off--diagonal forms.
Using distortion relations $\widehat{\mathbf{D}}=\nabla +\widehat{\mathbf{Z}}%
[\widehat{\mathcal{T}}]$ we can recover configurations with the Levi--Civita (LC) connection $%
\nabla $ which together with the distorting tensor $\widehat{\mathbf{Z}},$
and $\widehat{\mathbf{D}}$ are completely defined by $\widehat{\mathcal{T}}$
(for such models, by $(\mathbf{g},\mathbf{N})).$ Having constructed integral
varieties of some gravitational field equations in terms of $\widehat{%
\mathbf{D}},$ we can impose additional nonholonomic (non--integrable
constraints) when $\widehat{\mathbf{D}}_{\mid \widehat{\mathcal{T}}%
=0}\rightarrow \nabla $ and $\widehat{R}\rightarrow R,$ where $R$ is the
scalar curvature of $\nabla .$ As a result, it is possible to extract
generic off--diagonal solutions in GR and/or other theories with $\nabla .$

All terms in actions (\ref{act}) are stated by the same metric structure $%
\mathbf{g}=\{g_{\alpha \beta }\}$ for standard and/or modified models of
gravity theory generated by different (effective) Lagrangians when curvature
scalars $(\widehat{R},\widetilde{R},R,\breve{R}),$ for necessary type linear
connections, effective cosmological constant $\widehat{\Lambda },$
non--standard coupling of Ricci, $R_{\mu \nu },$ and energy--momentum, $%
T^{\mu \nu },$ tensors etc. A geometric model with $\ ^{[1]}\mathcal{L}$ can
be used for constructing general classes of generic off--diagonal solutions
in GR and modifications (including massive gravity). Such a theory models
effects with broken Lorentz invariance, non--standard effective anistoropic
fluid coupling and behavior of the polarized propagator in the
ultraviolet/infrared region. \ The solutions for $\ ^{[2]}\mathcal{L}$ will
be used as "bridges" between generating functions determining certain
classes of generic off--diagonal Einstein manifolds and effective models
with anisotropies, massive effects and resulting violations for Lorentz
symmetry. The effective Lagrange density $\ ^{[3]}\mathcal{L}$ is similar to
that a covariant renormalization gravity as in \cite{odints1}. For
corresponding conditions on $f,$ the value $\ ^{[4]}\mathcal{L}$ is related
to modified theories. In almost K\"{a}hler variables, theories of type $\
^{[1]}\mathcal{L}$--$\ ^{[3]}\mathcal{L}$ can be quantized using
non--perturbative methods for deformation quantization and A--brane
quantization, or as perturbative gauge like models \cite{vquant}.

In this paper, we study modified gravity models when the spacetime metrics for theories
derived from $\ ^{[3]}\mathcal{L}$ and/or $\ ^{[4]}\mathcal{L}$ are
approximated $~^{\diamond }\mathbf{g}_{\alpha \beta }=\eta _{\alpha \beta
}+h_{\alpha \beta }$ for  perturbative models with a
flat background metric $\eta _{\alpha \beta }.$ Such solutions are with massive
gravity effects, breaking of Lorentz symmetry and, for well stated
conditions, result into effective models with covariant renormalization.
Using frame transforms $\mathbf{g}_{\alpha \beta }=e_{~\alpha }^{\alpha
^{\prime }}e_{~\beta }^{\beta ^{\prime }}\ ^{\diamond }\mathbf{g}_{\alpha
^{\prime }\beta ^{\prime }},$ we shall connect the classical and quantum
perturbative solutions to the theories (and their nonlinear/generalized
solutions) determined by $\ ^{[1]}\mathcal{L}$, or $\ ^{[2]}\mathcal{L}$ and
background configurations for corresponding metric and connection structure.
Via nonholonomic constraints, the physical effects are derived to be similar
to those for some (super-) renormalizable theories $\ ^{[3]}\mathcal{L}$
and/or $\ ^{[4]}\mathcal{L}.$

\subsection{Nonholonomic distributions and noncommutative uncertainty
relations}

A class of (quantum type) modified noncommutative gravity theories are physically
motivated by Schr\"{o}dinger type uncertainty relations $\hat{u}^{\alpha }%
\hat{p}^{\beta }-\hat{p}^{\beta }\hat{u}^{\alpha }=i\hbar \widehat{\theta }%
^{\alpha \beta },$ where $\hbar $ is the Planck constant, $i^{2}=-1,$ and $%
\hat{u}^{\alpha }$ and $\hat{p}^{\beta }$ are, respectively, certain
coordinate and momentum type operators. Noncommutative
geometry/ gravity models are elaborated on "$\theta $--extensions" of tangent
Lorentz bundles $T\mathbf{V\rightarrow }$ $^{\theta }T\mathbf{V},$ (there
are used also co--tangent bundles $T^{\ast }\mathbf{V\rightarrow \ }$ $%
^{\theta }T^{\ast }\mathbf{V}$ etc). Such extensions are determined by
noncommutative complex distributions stated by "generalized uncertainty"
relations
\begin{equation}
u^{\alpha _{s}}u^{\beta _{s}}-u^{\beta _{s}}u^{\alpha _{s}}=i\theta ^{\alpha
_{s}\beta _{s}},  \label{ncom}
\end{equation}%
where the antisymmetric matrix $\theta =(\theta ^{\alpha _{s}\beta _{s}})$
can be taken with constant coefficients with respect to certain frame of
reference on "small" $\theta $--deformations $\mathbf{V\rightarrow \
^{\theta }V.}$ The label "$s$" is considered for three two dimensional (2-d)
"shells" $s=0,1,2,$ for a conventional splitting on $^{\theta }T\mathbf{V,}$
$\dim (\ ^{\theta }T\mathbf{V})=4+2s=2+2+2+2=8.$ The coordinates and
indices are respectively parameterized $u^{\alpha
_{s}}=(x^{i_{s}},y^{a_{s}}).$ We write $u^{\alpha }=(x^{i},y^{a})$ \ and
consider for {\small
\begin{eqnarray}
s &=&0:u^{\alpha _{0}}=(x^{i_{0}},y^{a_{0}})=u^{\alpha }=(x^{i},y^{a});\
\notag \\
s &=&1:u^{\alpha _{1}}=(x^{\alpha }=u^{\alpha
},y^{a_{1}})=(x^{i},y^{a},y^{a_{1}});  \notag \\
\ s &=&2:u^{\alpha _{2}}=(x^{\alpha _{1}}=u^{\alpha
_{1}},y^{a_{2}})=(x^{i},y^{a},y^{a_{1}},y^{a_{2}});  \label{coordparam} \\
\ s &=&3:u^{\alpha _{3}}=(x^{\alpha _{2}}=u^{\alpha
_{2}},y^{a_{3}})=(x^{i},y^{a},y^{a_{1}},y^{a_{2}},y^{a_{3}}),...,  \notag
\end{eqnarray}%
} when indices run corresponding values $%
i,j,...=1,2;a,b,...=3,4;a_{1},b_{1}...=5,6;a_{2},b_{2}...=7,8;$ $%
a_{3},b_{3}...=9,10,...$ and, for instance, $i_{1},j_{1},...=1,2,3,4;$ $%
i_{2}, j_{2},... =1,2,3,4,5,6;\ i_{3},j_{3},...=1,2,3,4,5,6,7,8;...$ In
brief, we shall write $u=(x,y); \ ^{1}u=(u,\ ^{1}y)=(x,y,\ ^{1}y),\ ^{2}u=(\
^{1}u,\ ^{2}y)=(x,y,\ ^{1}y,\ ^{2}y),...$

In this work, the geometric objects with $s>0$ are some $\theta $%
--deformations determined by distributions of type (\ref{ncom}) and
additional assumptions on "inner products", N--adapted connections etc. We
shall follow a geometric principle that for elaborating noncommutative
(quantum) geometric models there are considered classical nonlinear
configurations (exact solutions) for a gravity theory (\ref{act}) and then
elaborated  "small" deformations to noncommutative gravity models with
additional quantum parameters encoded for $s\neq 0.$

The noncommutative relations (\ref{ncom}) and coordinate parameterizations (%
\ref{coordparam}) are adapted to Whitney sums\footnote{%
in certain geometric and physical theories \cite{vexsolbranes,vnonc}, it is
used the term nonlinear connection, N--connection} $\ ^{s}\mathbf{N}:T\ ^{s}%
\mathbf{V}=h\mathbf{V}\oplus v\mathbf{V}\oplus \ ^{1}v\mathbf{V}\oplus \
^{2}v\mathbf{V.}$ This prescribes a local fibered structure on $\mathbf{\ }%
^{\theta }\mathbf{V,}$ when the coefficients of N--connection, $%
N_{i_{s}}^{a_{s}},$ for $\ ^{s}\mathbf{N}=N_{i_{s}}^{a_{s}}(\ ^{s}u,\theta
)dx^{i_{s}}\otimes \partial /\partial y^{a_{s}},$ and states a system of
N--adapted local bases with N-elongated partial derivatives, $\mathbf{e}%
_{\nu _{s}}=(\mathbf{e}_{i_{s}},e_{a_{s}}),$ and cobases with N--adapted
differentials, $\mathbf{e}^{\mu _{s}}=(e^{i_{s}},\mathbf{e}^{a_{s}}).$ For a
4-d commutative Lorentz manifold $\mathbf{V}$,
\begin{eqnarray}
&&\mathbf{e}_{i}=\frac{\partial }{\partial x^{i}}-\ N_{i}^{a}\frac{\partial
}{\partial y^{a}},\ e_{a}=\frac{\partial }{\partial y^{a}},  \label{nader} \\
&&e^{i}=dx^{i},\mathbf{e}^{a}=dy^{a}+\ N_{i}^{a}dx^{i},  \label{nadif}
\end{eqnarray}%
and on $s=1,2$ shells, (if any $\theta ^{\alpha _{s}\beta _{s}}\neq 0,$ the
coefficients depend on coordinates and such parameters, for instance, $%
N_{i_{s}}^{a_{s}}(\ ^{1}u,\ ^{2}y,\theta )$),
\begin{eqnarray}
\mathbf{e}_{i_{s}} &=&\frac{\partial }{\partial x^{i_{s}}}-\
N_{i_{s}}^{a_{s}}\frac{\partial }{\partial y^{a_{s}}},\ e_{a_{s}}=\frac{%
\partial }{\partial y^{a_{s}}},  \label{naders} \\
e^{i_{s}} &=&dx^{i_{s}},\mathbf{e}^{a_{s}}=dy^{a_{s}}+\
N_{i_{s}}^{a_{s}}dx^{i_{s}}.  \label{nadifs}
\end{eqnarray}%
The N--adapted operators (\ref{nader}) and (\ref{naders}) satisfy certain
anholonomy relations
\begin{equation}
\lbrack \mathbf{e}_{\alpha _{s}},\mathbf{e}_{\beta _{s}}]=\mathbf{e}_{\alpha
_{s}}\mathbf{e}_{\beta _{s}}-\mathbf{e}_{\beta _{s}}\mathbf{e}_{\alpha
_{s}}=W_{\alpha _{s}\beta _{s}}^{\gamma _{s}}\mathbf{e}_{\gamma _{s}},
\label{anhrel1}
\end{equation}%
completely defined by the N--connection coefficients and their partial
derivatives, $W_{i_{s}a_{s}}^{b_{s}}=\partial _{a_{s}}N_{i_{s}}^{b_{s}}$ and
$W_{j_{s}i_{s}}^{a_{s}}=\Omega _{i_{s}j_{s}}^{a_{s}},$ where the curvature
of N--connection is $\Omega _{i_{s}j_{s}}^{a_{s}}=\mathbf{e}_{j_{s}}\left(
N_{i_{s}}^{a_{s}}\right) -\mathbf{e}_{i_{s}}\left( N_{j_{s}}^{a_{s}}\right)
. $ The quantum noncommutative structure is encoded into such nonholonomic
distributions. The geometric objects with coefficients defined with respect
to N--adapted frames are called respectively distinguished metrics,
distinguished tensors etc (in brief, d--metrics, d--tensors etc)

Any metric structure $\ ^{s}\mathbf{g=\{g}_{\alpha _{s}\beta _{s}}\mathbf{\}}
$ on $\ ^{\theta }\mathbf{V}$ with $\theta =(\theta ^{\alpha _{s}\beta
_{s}}) $ can be written as a distinguished metric (d--metric)
\begin{eqnarray}
\ \ ^{s}\mathbf{g} &=&\ g_{i_{s}j_{s}}(\ ^{s}u,\theta )\ e^{i_{s}}\otimes
e^{j_{s}}+\ g_{a_{s}b_{s}}(\ ^{s}u,\theta )\mathbf{e}^{a_{s}}\otimes \mathbf{%
e}^{b_{s}}  \label{dm1s} \\
&=&g_{ij}(x)\ e^{i}\otimes e^{j}+g_{ab}(u)\ \mathbf{e}^{a}\otimes \mathbf{e}%
^{b}+  \notag \\
&&g_{a_{1}b_{1}}(\ ^{1}u,\theta )\ \mathbf{e}^{a_{1}}\otimes \mathbf{e}%
^{b_{1}}+\ g_{a_{2}b_{2}}(\ ^{2}u,\theta )\mathbf{e}^{a_{2}}\otimes \mathbf{e%
}^{b_{2}}.  \notag
\end{eqnarray}%
In coordinate frames, (\ref{dm1s}) is parameterized equivalently by generic
off--diagonal matrices (which can not be diagonalized via coordinate
transforms),
\begin{equation}
\ ^{s}\mathbf{g=}g_{\alpha _{s}\beta _{s}}e^{\alpha _{s}}\otimes e^{\beta
_{s}}=g_{\underline{\alpha }_{s}\underline{\beta }_{s}}du^{\underline{\alpha
}_{s}}\otimes du^{\underline{\beta }_{s}},\ s=0,1,2,...,  \label{metr}
\end{equation}%
where coefficients transform and $g_{\alpha _{s}\beta _{s}}=e_{\ \alpha
_{s}}^{\underline{\alpha }_{s}}e_{\ \beta _{s}}^{\underline{\beta }_{s}}g_{%
\underline{\alpha }_{s}\underline{\beta }_{s}},$ for respective frames and
local coordinate bases $e_{\alpha _{s}}=e_{\ \alpha _{s}}^{\underline{\alpha
}_{s}}(\ ^{s}u)\partial /\partial u^{\underline{\alpha }_{s}},\partial
_{\beta _{s}}:=\partial /\partial u^{\beta _{s}},$ when
\begin{equation*}
\ \ \underline{g}_{\alpha _{s}\beta _{s}}\left( \ ^{s}u\right) =\left[
\begin{array}{cc}
\ g_{i_{s}j_{s}}+\ h_{a_{s}b_{s}}N_{i_{s}}^{a_{s}}N_{j_{s}}^{b_{s}} &
h_{a_{s}e_{s}}N_{j_{s}}^{e_{s}} \\
\ h_{b_{s}e_{s}}N_{i_{s}}^{e_{s}} & \ h_{a_{s}b_{s}}%
\end{array}%
\right] ,s=0,1,2.
\end{equation*}

The metrics (\ref{dm1s}) and/or (\ref{metr}) encode higher shells
dependencies on noncommutative parameters $\theta =(\theta ^{\alpha
_{s}\beta _{s}})$ and distributions (\ref{ncom}). A self--consistent
approach to such theories is based on the Groenewold--Moyal product (star
product, or $\star $--product) inspired by the foundations of quantum
mechanics.  We apply a formalism elaborated in \cite{vasil} but modified
for nonholonomic distributions and connections with $\nabla \rightarrow
\mathbf{D}=\mathbf{\tilde{D},}$ or $=$ $\widehat{\mathbf{D}},$ and almost K%
\"{a}hler variables determined naturally by data $(\mathbf{g,N}),$ see \cite%
{vnonc}. The constructions are based on formal power series $C^{\infty }(%
\mathbf{V})[[\ell ]]$ with deformation parameter $\ell =i\hbar ,$ where $%
i^{2}=-1$ and where $\hbar =h/2\pi $ is used for respective Plank constants.
The symbol $\mathbf{\tilde{D}}$ is used for the Cartan connection which is
adapted by a canonical $N$--connection structure $\mathbf{\tilde{N}}$
completely determined by regular effective Lagrange generating function $%
\mathcal{L}(x,y).$ Such a value can be always prescribed on a (pseudo)
Riemannian spacetime $\mathbf{V}$ and modelled on $T\mathbf{V}$ following
the principle that the semi--spray equations are equivalent to the
Euler--Lagrange ones, see details in \cite{vfinsl}.

There are induced by any $\mathcal{L}(x,y)$ other important geometric
structures. We remember that an almost complex structure is a linear operator
$\mathbf{J}$ acting on vectors on $T\mathbf{V}$ via actions on N--adapted
frames, $\mathbf{J}(\mathbf{e}_{i})=-e_{2+i}$ and $\mathbf{J}(e_{2+i})=%
\mathbf{e}_{i},$ where $\mathbf{J\circ J=-}\mathbb{I}\mathbf{,}$ for $%
\mathbb{I}$ being the unity matrix. If such a structure is canonical, we can
write $\mathbf{\tilde{J}}$ for $\mathbf{N}=\mathbf{\tilde{N}}.$ A canonical
almost K\"{a}hler space with $\mathbf{g}=\mathbf{\tilde{g},}$ $\mathbf{N=%
\tilde{N}}$ and $\mathbf{J=\tilde{J}}$ canonically defined by $\mathcal{L},$
when $\mathbf{\tilde{\theta}(\cdot ,\cdot )}:=\mathbf{\tilde{g}}\left(
\mathbf{\tilde{J}\cdot ,\cdot }\right) .$ We can introduce $\mathbf{\tilde{%
\theta}}=d\tilde{\omega}$ for $\tilde{\omega}:=\frac{1}{2}\frac{\partial
\mathcal{L}}{\partial y^{i}}dx^{i},$ where $d\mathbf{\tilde{\theta}}=dd%
\tilde{\omega}=0.$ If $\theta $ is  related to a canonical $%
\mathbf{\tilde{\theta}}$ via frame transforms, $\mathbf{\theta }_{\alpha
^{\prime }\beta ^{\prime }}e_{\ \alpha }^{\alpha ^{\prime }}e_{\ \beta
}^{\beta ^{\prime }}=\mathbf{\tilde{\theta}}_{\alpha \beta },$ we positively
construct an almost K\"{a}hler structure. Fixing a convenient $\mathcal{L}$
(with a corresponding N--splitting $\mathbf{\tilde{N}),}$ we generate
equivalent geometric models of nonholonomic manifolds with $(\mathbf{g,N}%
)\approx (\mathbf{\tilde{g},\tilde{N}})\approx (\mathbf{\theta ,J})\approx (%
\mathbf{\tilde{\theta},\tilde{J}}).$ The models with shell d--metric, $\ ^{s}%
\mathbf{g,}$ and almost symplectic, $\ ^{s}\mathbf{\theta =\{}\theta ^{\mu
_{s}\nu _{s}}\mathbf{\},}$ are convenient for constructing exact solutions
and/or study Finsler--Lagrange geometries but the almost symplectic ones can
be used for noncommutative geometry with distributions of type (\ref{ncom})
and/or deformation quantization. For simplicity, we shall omit the left
label "s" if that will not result in ambiguities.

The canonical (Cartan) covariant star product $^{\theta }T\mathbf{V}$ is
introduced as
\begin{equation}
\alpha \tilde{\star}\beta :=\sum\limits_{k}\frac{\ell ^{k}}{k!}\mathbf{%
\theta }^{\mu _{1}\nu _{1}}...\mathbf{\theta }^{\mu _{k}\nu _{k}}(\mathbf{D}%
_{\mu _{1}}\ldots \mathbf{D}_{\mu _{k}})\cdot (\mathbf{D}_{\nu _{1}}\ldots
\mathbf{D}_{\nu _{k}}),  \label{cstp}
\end{equation}%
for $\mathbf{D}=\mathbf{\tilde{D}}$ (or $=\widehat{\mathbf{D}}$). The
product $\tilde{\star}$ is adapted to a N--connection structure (\ref%
{whitney}) and maps d--tensors into d--tensors. For $\mathbf{D}\rightarrow
\nabla ,$ this operator transforms into similar noncommutative
generalizations of the (pseudo) Riemann geometry if $\mathbf{\theta }$ is
fixed for a symplectic manifold, $\tilde{\star}\rightarrow \star$. It is
possible to define  $s$--shell associative star operators $\ ^{s}\tilde{\star}$\ if $\mathbf{D%
}_{\mu _{s}}=(\mathbf{D}_{i_{s}},\mathbf{D}_{a_{s}}),$
\begin{equation*}
\alpha \ (\ ^{s}\tilde{\star})\beta =\sum\limits_{k}\frac{\ell ^{k}}{k!}%
\mathbf{\theta }^{a_{1}b_{1}}...\mathbf{\theta }^{a_{k}b_{k}}(\mathbf{D}%
_{a_{1}}\ldots \mathbf{D}_{a_{k}})\cdot (\mathbf{D}_{b_{1}}\ldots \mathbf{D}%
_{b_{k}}).
\end{equation*}%
The star product (\ref{cstp}) can be re--expressed in the form
\begin{equation}
\alpha \tilde{\star}\beta :=\alpha \beta +\sum\limits_{k}^{\infty }\ell ^{k}%
\mathbf{C}_{k}(\alpha ,\beta ),  \label{productapr}
\end{equation}%
where the bilinear operators $\mathbf{C}_{k}$ are N--adapted, i.e.
d--operators.

The product $\tilde{\star}$ (\ref{productapr}) satisfies such properties:

\begin{enumerate}
\item associativity, $\alpha \tilde{\star}(\beta \tilde{\star}\gamma
)=(\alpha \tilde{\star}\beta )\tilde{\star}\gamma ;$

\item it is defined the Poisson bracket, $\mathbf{C}_{1}(\alpha ,\beta
)=\{\alpha ,\beta \}=\theta ^{\mu _{s}\nu _{s}}\mathbf{D}_{\mu _{s}}\alpha
\cdot \mathbf{D}_{\nu _{s}}\beta ,$ distorted by the nonholonomically
induced torsion completely defined by the metric (almost symplectic)
structure; here we note antisymmetry, $\{\alpha ,\beta \}=-\{\beta ,\alpha
\},$ and the Jacoby identity,
\begin{equation*}
\{\alpha ,\{\beta ,\gamma \}\}+\{\gamma ,\{\alpha ,\beta \}\}+\{\beta
,\{\alpha ,\gamma \}\}=0;
\end{equation*}

\item there is an N--adapted stability of type $\alpha \tilde{\star}\beta
=\alpha \cdot \beta $ if $\ ^{s}\mathbf{D}\alpha =0$ or $\ ^{s}\mathbf{D}%
\beta =0;$

\item the Moyal symmetry, $\mathbf{C}_{k}(\alpha ,\beta )=(-1)^{k}\mathbf{C}%
_{k}(\beta ,\alpha );$

\item the N-adapted derivation with Leibniz rule,{\small \ }%
\begin{eqnarray*}
\ ^{s}\mathbf{D(}\alpha \tilde{\star}\beta ) &=&(\ ^{s}\mathbf{D}\alpha )%
\tilde{\star}\beta +\alpha \tilde{\star}(\ ^{s}\mathbf{D}\beta ) \\
&=&\left( (h\mathbf{D+}\ ^{s}v\mathbf{D)}\alpha \right) \tilde{\star}\beta
+\alpha \tilde{\star}((h\mathbf{D+}\ ^{s}v\mathbf{D)}\beta ).
\end{eqnarray*}%
For applications in quantum physics, it is important the Hermitian property,
$\overline{\alpha \tilde{\star}\beta }=\overline{\beta }\tilde{\star}%
\overline{\alpha },$ with complex conjugation "-", when
 $
\mathbf{g}_{\alpha _{s}\beta _{s}}=\frac{1}{2}\left( \overline{\mathbf{e}}%
_{\alpha _{s}}\tilde{\star}\mathbf{e}_{\beta _{s}}+\overline{\mathbf{e}}%
_{\beta _{s}}\tilde{\star}\mathbf{e}_{\alpha _{s}}\right)$.
It should be noted that because $\ ^{s}\mathbf{\tilde{D}}\tilde{\theta}=0,$
we can write $\mathbf{\theta }^{\mu _{s}\nu _{s}}\tilde{\star}\chi =\mathbf{%
\theta }^{\mu _{s}\nu _{s}}\cdot \chi .$
\end{enumerate}

Using canonical almost symplectic data $(\tilde{\star},\ ^{s}\mathbf{\tilde{D%
}),}$ it is possible to elaborate an associative star product calculus in
noncommutative spacetime which is completely defined by the metric structure
in N--adapted form and keeps the covariant property. We derive a N--adapted
local frame structure $\mathbf{\tilde{e}}_{\alpha _{s}}=(\mathbf{e}_{i_{s}},%
\mathbf{\tilde{e}}_{a_{s}})$ which can be constructed by noncommutative
deformations of $\mathbf{e}_{\alpha },$%
\begin{eqnarray}
\mathbf{\tilde{e}}_{\alpha _{s}\ }^{\ \underline{\alpha }_{s}} &=&\mathbf{e}%
_{\alpha _{s}\ }^{\ \underline{\alpha }_{s}}+i\theta ^{\gamma _{s}\beta _{s}}%
\mathbf{e}_{\alpha _{s}\ \gamma _{s}\beta _{s}}^{\ \underline{\alpha }%
_{s}}+\theta ^{\gamma _{s}\beta _{s}}\theta ^{\tau _{s}\mu _{s}}\mathbf{e}%
_{\alpha _{s}\ \gamma _{s}\tau _{s}\mu _{s}}^{\ \underline{\alpha }_{s}}+%
\mathcal{O}(\theta ^{3}),  \label{ncfd} \\
\mathbf{\tilde{e}}_{\ \star \underline{\alpha }_{s}}^{\alpha _{s}} &=&%
\mathbf{e}_{\alpha _{s}\ }^{\ \underline{\alpha }_{s}}+i\theta ^{\gamma
_{s}\beta _{s}}\mathbf{e}_{\ \underline{\alpha }_{s}\gamma _{s}\beta
_{s}}^{\alpha _{s}}+\theta ^{\gamma _{s}\beta _{s}}\theta ^{\tau _{s}\mu
_{s}}\mathbf{e}_{\ \underline{\alpha }_{s}\gamma _{s}\beta _{s}\tau _{s}\mu
_{s}}^{\alpha _{s}}+\mathcal{O}(\theta ^{3}),  \notag
\end{eqnarray}%
subjected to the condition $\mathbf{\tilde{e}}_{\ \star \underline{\alpha }%
_{s}}^{\alpha _{s}}\star \mathbf{\tilde{e}}_{\alpha _{s}\ }^{\ \underline{%
\beta }_{s}}\ =\delta _{\underline{\alpha }_{s}}^{\ \underline{\beta }_{s}},$
where $\delta _{\underline{\alpha }}^{\ \underline{\beta }}$ is the
Kronecker tensor. The values $\mathbf{e}_{\alpha _{s}\ \gamma _{s}\beta
_{s}}^{\ \underline{\alpha }_{s}}$ and $\mathbf{e}_{\alpha _{s}\ \gamma
_{s}\tau _{s}\mu _{s}}^{\ \underline{\alpha }_{s}}$ can be written in terms
of $\mathbf{e}_{\alpha _{s}\ }^{\ \underline{\alpha }_{s}},\theta ^{\gamma
_{s}\beta _{s}}$ and the spin distinguished connection corresponding to $\
^{s}\mathbf{\tilde{D},}$ or $\ ^{s}\widehat{\mathbf{D}},$ see similar
formulas in \cite{vnonc,vasil}.

The noncommutative deformations of a metric, $\mathbf{g}$ $\rightarrow \ ^{s}%
\mathbf{g,}$ can be defined and computed
 $\ ^{s}\mathbf{g}_{\alpha _{s}\beta _{s}}=\frac{1}{2}\eta _{\underline{\alpha
}_{s}\underline{\beta }_{s}}\left[ \mathbf{\tilde{e}}_{\alpha _{s}\ }^{\
\underline{\alpha }_{s}}\star \left( \mathbf{\tilde{e}}_{\beta _{s}\ }^{\
\underline{\beta }_{s}}\right) ^{+}+\mathbf{\tilde{e}}_{\beta _{s}\ }^{\
\underline{\beta }_{s}}\star \left( \mathbf{\tilde{e}}_{\alpha _{s}\ }^{\
\underline{\alpha }_{s}}\right) ^{+}\right]$,
where $\left( \ldots \right) ^{+}$ is used for the Hermitian conjugation and
$\eta _{\underline{\alpha }_{s}\underline{\beta }_{s}}$ denotes the flat
Minkowski spacetime metric extended on $T\mathbf{V.}$ We can parameterize
the noncommutative and nonholonomic transforms when $\ ^{s}\mathbf{g}%
_{\alpha _{s}\beta _{s}}(\ ^{s}u,\theta )$ (\ref{dm1s}) is with real
coefficients which for $s=1,2$ depend only on even powers of $\theta ,$%
\begin{eqnarray}
g_{i}(u) &=&\grave{g}_{i}(x^{k}),\ h_{a}=\grave{h}_{a}(u),\ h_{a_{s}}=\grave{%
h}_{a_{s}}(u)+\ ^{2}h_{a_{s}}(u)\theta ^{2}+\mathcal{O}(\theta ^{4}),  \notag
\\
N_{i_{s}}^{a_{s}}(\ ^{s}u,\theta ) &=&\grave{N}_{i_{s}}^{a_{s}}(\ ^{s}u)+\
^{2}N_{i_{s}}^{a_{s}}(\ ^{s}u)\theta ^{2}+\mathcal{O}(\theta ^{4}).
\label{coefm}
\end{eqnarray}%
This allows us to treat $\theta $ as some integration parameters related to
superposition of Killing symmetries and anholonomic frame transformations
\cite{geroch}.  For simplicity, we shall not write $\theta $ in
explicit form if that will not result in ambiguities for any notation of type $N_{i_{s}}^{a_{s}}(\ ^{s}u,\theta
),N_{i_{s}}^{a_{s}}(\theta ),$ or $N_{i_{s}}^{a_{s}}(\ ^{s}u).$

\section{Decoupling \& Integration of (Non)Commutati\-ve Modifi\-ed Massive
Gravity}

\label{sec3}

\subsection{Effective Einstein equations}

Applying a N--adapted variational calculus for $\ ^{[1]}\mathcal{L}$ on $%
\mathbf{V,}$ see details in \cite{vexsolbranes,rhamhr,odints}, we derive the
equations of motion for  4--d modified massive gravity
\begin{equation}
(\partial _{\widehat{R}}\widehat{f})\widehat{\mathbf{R}}_{\alpha \beta }-%
\frac{1}{2}\widehat{f}(\widehat{R})\mathbf{g}_{\alpha \beta }+\mathring{\mu}%
^{2}\mathbf{X}_{\alpha \beta }=M_{Pl}^{-2}\mathbf{T}_{\alpha \beta },
\label{mgrfe}
\end{equation}%
where $M_{Pl}$ is the Plank mass, $\widehat{\mathbf{R}}_{\mu \nu }$ is the
Einstein tensor for a pseudo--Riemannian metric $\mathbf{g}_{\mu \nu }$ and $%
\widehat{\mathbf{D}},$ $\mathbf{T}_{\mu \nu }$ is the standard
energy--momentum tensor for matter. The effective energy--momentum tensor $%
\mathbf{X}_{\mu \nu }$ is determined by the potential of graviton $\mathcal{U%
}=\mathcal{U}_{2}+\alpha _{3}\mathcal{U}_{3}+\alpha _{4}\mathcal{U}_{4},$
where $\alpha _{3}$ and $\alpha _{4}$ are free parameters and $\mathcal{U}%
_{2},\mathcal{U}_{3}$ and $\mathcal{U}_{4}$ are certain polynomials on
traces of some other polynomials of the matrix $\mathcal{K}_{\mu }^{\nu
}=\delta _{\mu }^{\nu }-\left( \sqrt{g^{-1}\Sigma }\right) _{\mu }^{\nu }.$
There are involved four St\"{u}ckelberg fields $\phi ^{\underline{\mu }}$ as
$\ \Sigma _{\mu \nu }=\partial _{\mu }\phi ^{\underline{\mu }}\partial _{\nu
}\phi ^{\underline{\nu }}\eta _{\underline{\mu }\underline{\nu }},$ when $%
\eta _{\underline{\mu }\underline{\nu }}=(1,1,1,-1).$ A parameter choice $%
\alpha _{3}=(\alpha -1)/3,\alpha _{4}=(\alpha ^{2}-\alpha +1)/12$ is optimal if we want
to avoiding potential ghost instabilities and $\mathbf{X}_{\mu \nu }=\alpha
^{-1}\mathbf{g}_{\mu \nu }$. It is  possible to find exact off--diagonal
solutions of (\ref{mgrfe}) if we fix the coefficients $\{N_{i}^{a}\}$ of $%
\mathbf{N}$ and local frames for $\widehat{\mathbf{D}}$ when $\widehat{R}%
=const$ and $\partial _{\alpha }\widehat{f}(\widehat{R})=(\partial _{%
\widehat{R}}\widehat{f})\times \partial _{\alpha }\widehat{R}=0.$ In
general, $\partial _{\alpha }R\neq 0$ and $\partial _{\alpha }f(\breve{R}%
)\neq 0.$ For simplicity, we shall consider configurations withe energy
momentum sources $\mathbf{T}_{\mu \nu }$ and effective $\mathbf{X}_{\mu \nu
} $ which (using frame transforms) can be parameterized with respect to
N--adapted frames (\ref{nader}) and (\ref{nadif}) in the form {\small
\begin{equation}
\Upsilon _{\beta }^{\alpha }=\frac{1}{M_{Pl}^{2}(\partial _{\widehat{R}}%
\widehat{f})}(\mathbf{T}_{\beta }^{\alpha }+\alpha ^{-1}\mathbf{X}_{\beta
}^{\alpha })=\frac{1}{M_{Pl}^{2}(\partial _{\widehat{R}}\widehat{f})}(\
^{m}T+\alpha ^{-1})\delta _{\beta }^{\alpha }=(\hat{\Upsilon}+\ \mathring{%
\Upsilon})\delta _{\beta }^{\alpha },  \label{effectsourc}
\end{equation}%
} for constant values $\ \hat{\Upsilon}:=M_{Pl}^{-2}(\partial _{\widehat{R}}%
\widehat{f})^{-1}\ ^{m}T$ and $\ \mathring{\Upsilon}=M_{Pl}^{-2}(\partial _{%
\widehat{R}}\widehat{f})^{-1}\alpha ^{-1}.$

All above constructions can be extended in a (non) commutative form to extra
shells $s=1,2,...$ via formal re--definition of indices for higher
dimensions, adapting with respect to bases (\ref{naders}) and (\ref{nadifs})
generated by noncommutative nonholonomic distributions (\ref{ncom}). Under
very general assumptions, the effective source can be parameterized in the
form
\begin{equation}
\mathbf{\Upsilon }_{~\delta _{s}}^{\beta _{s}}=(\ ^{s}\hat{\Upsilon}+\ \ ^{s}%
\mathring{\Upsilon})\mathbf{\delta }_{~\delta _{s}}^{\beta _{s}}.
\label{source1b}
\end{equation}%
The sources with $s=1$ and $2$ are considered as certain effective ones on
"fibers" with noncommutative variables. For the simplest models, they can
associated to certain effective cosmological constants or stated to be zero.
Geometrically, such sources can be defined as N--adapted lifts from the base
to the total space. Quantum corrections to certain classes of solutions can
be modelled via certain stochastic (Finsler like or generalized ones \cite%
{vfinsl,vquant}) corrections to generating functions and sources. Such $(\
^{s}\hat{\Upsilon}+\ \ ^{s}\mathring{\Upsilon})$--terms encode via
nonholonomic constraints and the canonical d--connection $\ ^{s}\widehat{%
\mathbf{D}}$ various physically important information on possible
modifications of the GR theory on certain Lorentz manifolds backgrounds and
generalized (co) tangent commutative and/or noncommutative bundles. On $%
^{\theta }T\mathbf{V,}$ modified gravitational field equations of type (\ref%
{mgrfe}) are written
\begin{eqnarray}
&&\ ^{s}\widehat{\mathbf{R}}_{\ \beta _{s}\delta _{s}}-\frac{1}{2}\mathbf{g}%
_{\beta _{s}\delta _{s}}\ ^{sc}\widehat{R}=\mathbf{\Upsilon }_{\beta
_{s}\delta _{s}},  \label{cdeinst} \\
&&\widehat{L}_{a_{s}j_{s}}^{c_{s}}=e_{a_{s}}(N_{j_{s}}^{c_{s}}),\ \widehat{C}%
_{j_{s}b_{s}}^{i_{s}}=0,\ \Omega _{\ j_{s}i_{s}}^{a_{s}}=0,  \label{lcconstr}
\end{eqnarray}%
where sources $\mathbf{\Upsilon }_{\beta _{s}\delta _{s}}$ (\ref{source1b})
for $s=0$ are formally defined as in GR but on (non) commutative fibers
using (\ref{cstp}) and (\ref{productapr}). For commutative extra dimensions,
we can take n $\mathbf{\Upsilon }_{\beta _{s}\delta _{s}}\rightarrow
\varkappa T_{\beta _{s}\delta _{s}}$ \ for $\ ^{s}\widehat{\mathbf{D}}%
\rightarrow \ ^{s}\nabla ,$ with effective sources as canonical lifts from
the base 4--d spacetime determined by distributions (\ref{ncom}). In
general, the solutions of (\ref{cdeinst}) are with nonholonomically induced
torsion. If the conditions (\ref{lcconstr}) are satisfied, the nonholonomically induced torsion is constrained to be zero
and we get the Levi--Civita, LC, connection.

There are two classes of (non) commutative theories with effective
gravitational field equations of type \ (\ref{cdeinst}) and (possible)
constraints (\ref{lcconstr}):
\begin{enumerate}
\item Models generated for $s\neq 0$ as "twisted" noncommutative products
adapted to nonholonomic (complex and/or real) distributions. For $s=0,$ the
equations (\ref{cdeinst}) are equivalent to (\ref{mgrfe}). To derive such
equations we can use any action (\ref{act}) and elaborate corresponding
massive, modified Einstein gravity theories with generalized
connections.

\item There is a more general class of theories when (\ref{cdeinst}) and (%
\ref{source1b}) are constructed directly on $T\ ^{s}\mathbf{V}$ and/or $%
^{\theta }T\mathbf{V}$ with arbitrary N--connections. We shall not study
such theories in this works even, for instance, certain classes of solutions
compactified / warped / trapped on base commutative spacetimes may play an
important role in modified gravity theories \ (see, for instance, Finsler
branes etc \cite{vfinsl}).
\end{enumerate}

\subsection{N--adapted symmetries}

The decoupling property of N--adapted equations (\ref{cdeinst}) can be
proved in a straightforward form for $\ ^{s}\widehat{\mathbf{D}}$ and metrics of
type
{\small
\begin{eqnarray}
\ ^{s}\mathbf{g} &=&\ g_{i}(x^{k})dx^{i}\otimes dx^{i}+h_{a}(x^{k},y^{4})%
\mathbf{e}^{a}\otimes \mathbf{e}^{b}+  \label{ansk} \\
&&h_{a_{1}}(u^{\alpha },y^{6},\theta )\ \mathbf{e}^{a_{1}}\otimes \mathbf{e}%
^{a_{1}}+h_{a_{2}}(u^{\alpha _{1}},y^{8},\theta )\ \mathbf{e}^{a_{2}}\otimes
\mathbf{e}^{b_{2}},  \notag
\end{eqnarray}%
\begin{eqnarray}
&&\mbox{where }\ \mathbf{e}^{a}=dy^{a}+N_{i}^{a}dx^{i},\mbox{\ for \ }%
N_{i}^{3}=n_{i}(x^{k},y^{4}),N_{i}^{4}=w_{i}(x^{k},y^{4});  \label{ncon} \\
&&\mathbf{e}^{a_{1}}=dy^{a_{1}}+N_{\alpha }^{a_{1}}du^{\alpha },%
\mbox{\ for
\ }N_{\alpha }^{5}=\ ^{1}n_{\alpha }(u^{\beta },y^{6},\theta ),N_{\alpha
}^{6}=\ ^{1}w_{\alpha }(u^{\beta },y^{6},\theta );  \notag \\
&&\mathbf{e}^{a_{2}}=dy^{a_{2}}+N_{\alpha _{1}}^{a_{2}}du^{\alpha _{1}},%
\mbox{\ for \ }N_{\alpha _{1}}^{7}=\ ^{2}n_{\alpha _{1}}(u^{\beta
_{1}},y^{8},\theta ),N_{\alpha _{1}}^{8}=\ ^{2}w_{\alpha }(u^{\beta
_{1}},y^{8},\theta ).  \notag
\end{eqnarray}%
} Such ansatz contains a Killing vector $\partial /\partial y^{7}$ because
the coordinate $y^{7}$ is not contained in the coefficients of such metrics.
If $\theta \rightarrow 0,$ we generate off--diagonal metrics of 4--d
effective Einstein equations with Killing vector $\partial /\partial y^{3}$
(it is possible to construct solutions with non--Killing symmetries, see
details in \cite{vexsolbranes,vfinsl}).

Let us define the values
{\footnotesize
\begin{eqnarray}
\phi =\ln |\frac{\partial _{4}h_{3}}{\sqrt{|h_{3}h_{4}|}}|,\ \gamma :=\partial _{4}\ln \frac{|h_{3}|^{3/2}}{|h_{4}|},\
 \alpha _{i}=\partial _{4}\sqrt{|h_{3}|}\partial _{i}\phi ,\ \beta
=\partial _{4}\sqrt{|h_{3}|}\partial _{4}\phi , \label{genfc} \\
\ ^{1}\phi =\ln | \frac{\partial _{6}h_{5}}{\sqrt{|h_{5}h_{6}|}} |,
\ ^{1}\gamma :=\partial _{6}\ln \frac{|h_{5}|^{3/2}}{|h_{6}|},
 \ ^{1}\alpha _{\tau }=\partial _{6}\sqrt{|h_{5}|}\partial _{\tau }\
^{1}\phi ,\ ^{1}\beta =\partial _{6}\sqrt{|h_{5}|}\partial _{\tau }\
^{1}\phi ,  \notag \\
\ ^{2}\phi =\ln | \frac{\partial _{8}h_{7}}{\sqrt{|h_{7}h_{8}|}}|,\ ^{2}\gamma :=\partial _{8}\ln \frac{|h_{7}|^{3/2}}{|h_{8}|},
  \ ^{2}\alpha _{\tau _{1}}=\partial _{8}\sqrt{|h_{7}|}\partial _{\tau
_{1}}\ ^{2}\phi ,\ ^{2}\beta =\partial _{8}\sqrt{|h_{7}|}\partial _{\tau
_{1}}\ ^{2}\phi , \notag
\end{eqnarray}%
}
where $\phi (x^{k},y^{4}),\ ^{1}\phi (u^{\alpha },y^{6},\theta )$ and $\
^{2}\phi (u^{\alpha _{1}},y^{8},\theta )$ will be used as generating
functions which can be re--defined by corresponding transforms of $%
h_{a},h_{a_{1}}$ and $h_{a_{2}}$ (or inversely). A tedious computation of
the N--adapted coefficients of the Ricci tensor of $\ ^{s}\widehat{\mathbf{D}%
}$ for the ansatz (\ref{ansk}) allows us to express (\ref{cdeinst}) as
\begin{eqnarray}
\widehat{R}_{1}^{1} &=&\widehat{R}_{2}^{2}=-\Lambda (x^{k}),\ \widehat{R}%
_{3}^{3}=\widehat{R}_{4}^{4}=-\ ^{v}\Lambda (x^{k},y^{4}),  \label{sourc1} \\
\widehat{R}_{5}^{5} &=&\widehat{R}_{6}^{6}=-\ _{1}^{v}\Lambda (u^{\beta
},y^{6}),\ \widehat{R}_{7}^{7}=\widehat{R}_{8}^{8}=-\ _{2}^{v}\Lambda
(u^{\beta _{1}},y^{8}),  \notag
\end{eqnarray}%
with nontrivial effective (polarized gravitational constants) $\Lambda $%
--sources related to $\mathbf{\Upsilon }_{\beta _{s}\delta _{s}}$ (\ref%
{source1b}) via formulas
\begin{eqnarray*}
\mathbf{\Upsilon }_{1}^{1} &=&\mathbf{\Upsilon }_{2}^{2}=\ ^{v}\Lambda +\
_{1}^{v}\Lambda +\ _{2}^{v}\Lambda ,\mathbf{\Upsilon }_{3}^{3}=\mathbf{%
\Upsilon }_{4}^{4}=\Lambda +\ _{1}^{v}\Lambda +\ _{2}^{v}\Lambda , \\
\mathbf{\Upsilon }_{5}^{5} &=&\mathbf{\Upsilon }_{6}^{6}=\Lambda +\
_{1}^{v}\Lambda +\ _{2}^{v}\Lambda ,\mathbf{\Upsilon }_{7}^{7}=\mathbf{%
\Upsilon }_{8}^{8}=\Lambda +\ ^{v}\Lambda +\ _{1}^{v}\Lambda .
\end{eqnarray*}%
For certain models of extra dimension gravity, and re--defining the
integration functions, we can put $\ _{1}^{v}\Lambda =\ _{2}^{v}\Lambda =\
^{\circ }\Lambda =const.$

Using formulas (\ref{sourc1}), we can compute the Ricci scalar $\ ^{sc}%
\widehat{R}:=\ ^{s}\widehat{\mathbf{R}}_{\ \ \beta _{s}}^{\beta _{s}},$ $\
^{sc}\widehat{R}=2(\widehat{R}_{1}^{1}+\widehat{R}_{3}^{3}+\widehat{R}%
_{5}^{5}).$ As a result, we prove that there are certain N--adapted
symmetries of the Einstein d--tensor $\widehat{E}_{\ \beta _{s}\delta
_{s}}:=\ ^{s}\widehat{\mathbf{R}}_{\ \beta _{s}\delta _{s}}-\frac{1}{2}%
\mathbf{g}_{\beta _{s}\delta _{s}}\ ^{sc}\widehat{R}$ for the ansatz (\ref%
{ansk}) with Killing vector $\partial /\partial y^{7}:$%
{\small
\begin{eqnarray*}
s &=&1:\widehat{E}_{1}^{1}=\widehat{E}_{2}^{2}=-(\widehat{R}_{3}^{3}+%
\widehat{R}_{5}^{5}),\widehat{E}_{3}^{3}=\widehat{E}_{4}^{4}=-(\widehat{R}%
_{1}^{1}+\widehat{R}_{5}^{5}), \\
&& \widehat{E}_{5}^{5}=\widehat{E}_{6}^{6}=-(\widehat{R}_{1}^{1}+\widehat{R}%
_{3}^{3}), \\
s &=&2:\ \widehat{E}_{1}^{1}=\widehat{E}_{2}^{2}=-(\widehat{R}_{3}^{3}+%
\widehat{R}_{5}^{5}+\widehat{R}_{7}^{7}),\widehat{E}_{3}^{3}=\widehat{E}%
_{4}^{4}=-(\widehat{R}_{1}^{1}+\widehat{R}_{5}^{5}+\widehat{R}_{7}^{7}), \\
&&\widehat{E}_{5}^{5} =\widehat{E}_{6}^{6}=-(\widehat{R}_{1}^{1}+\widehat{R}%
_{3}^{3}+\widehat{R}_{7}^{7}),\widehat{E}_{7}^{7}=\widehat{E}_{8}^{8}=-(%
\widehat{R}_{1}^{1}+\widehat{R}_{3}^{3}+\widehat{R}_{5}^{5}).
\end{eqnarray*}
}
Such symmetries of the linear connection $\ ^{s}\widehat{\mathbf{D}}$ which
can be nonholonomically constrained to $\nabla $ are important for
decoupling and constructing generic off--diagonal solutions of gravitational
field equations.

\subsection{Decoupling and integration of gravitational field eqs}

The ansatz (\ref{ansk}) for $\ g_{i}(x^{k})=\epsilon _{i}e^{\psi
(x^{k})},\epsilon _{i}=1,$ with nonzero $\partial _{4}\phi ,\partial
_{4}h_{a},$ $\partial _{6}\ ^{1}\phi ,\partial _{6}h_{a_{1}},\partial _{8}\
^{2}\phi ,\partial _{8}h_{a_{2}},...$ transforms (\ref{cdeinst}) into such a
system of PDEs:
\begin{eqnarray}
&&\epsilon _{1}\partial _{11}\psi +\epsilon _{2}\partial _{22}\psi =2\Lambda
(x^{k}),  \label{einst1h} \\
&&\partial _{4}\phi \ \partial _{4}h_{3}=2h_{3}h_{4}\ ^{v}\Lambda
(x^{k},y^{4}),  \label{einst1v} \\
&&\partial _{44}n_{i}+\gamma \partial _{4}n_{i}=0,\ \beta w_{i}-\alpha
_{i}=0,\   \notag \\
&&  \notag \\
&&\partial _{6}\ ^{1}\phi \ \partial _{6}h_{5}=2h_{5}h_{6}\ _{1}^{v}\Lambda
(u^{\beta },y^{6},\theta ),  \label{einst1v1} \\
&&\partial _{66}\ ^{1}n_{\tau }+\ ^{1}\gamma \partial _{6}\ ^{1}n_{\tau
}=0,\ ^{1}\beta \ ^{1}w_{\tau }-\ ^{1}\alpha _{\tau }=0,\   \notag \\
&&  \notag \\
&&\partial _{8}\ ^{2}\phi \ \partial _{6}h_{7}=2h_{7}h_{8}\ _{2}^{v}\Lambda
(u^{\beta _{1}},y^{8}, \theta),  \label{einst1v2} \\
&&\partial _{88}\ ^{2}n_{\tau _{1}}+\ ^{2}\gamma \partial _{8}\ ^{2}n_{\tau
_{1}}=0,\ ^{2}\beta \ ^{2}w_{\tau _{1}}-\ ^{2}\alpha _{\tau _{1}}=0.\
\notag
\end{eqnarray}%
Let us explain the decoupling property of (\ref{einst1h})--(\ref{einst1v2}%
):\ The equation in (\ref{einst1h}) is just the 2--de Laplace/Poisson
equation which can be solved for any given $\Lambda (x^{k}).$ The first
equation in the group (\ref{einst1v}) allows us to express $h_{3}$ through $%
h_{4}$ (or inversly) for any given generating function $\phi $ and
nontrivial source $\ ^{v}\Lambda ,$ see details in \cite{vexsolbranes,vfinsl}%
. Having $h_{3}$ and $h_{4},$ we can compute the coefficients $\gamma
,\alpha _{i},\beta ,$ in (\ref{genfc}) and find $n_{i}$ and $w_{i},$
respectively, integating two times on $y^{4}$ and solving an algebraic
equations. In a similar form, we can construct solutions for $h_{5}$ and $%
h_{6}$ and then for $\ ^{1}n_{\tau }$ and $\ ^{1}w_{\tau }$ for any
generating function $\ ^{1}\phi $ and source $\ _{1}^{v}\Lambda $ in the
group (\ref{einst1v1})$.$ The same procedured can be used for constructing
solutions for $h_{7}$ and $h_{7}$ and then for $\ ^{1}n_{\tau _{1}}$ and $\
^{1}w_{\tau _{1}}$ because the system (\ref{einst1v2}) is a similar
extension of (\ref{einst1v1}) on variable $y^{8}.$

We generate solutions of (\ref{cdeinst}) and LC--conditions (\ref{lcconstr})
by any {\small
\begin{eqnarray}
ds_{K}^{2} &=&\epsilon _{i}e^{\psi (x^{k},\mathring{\mu})}(dx^{i})^{2}+\frac{%
\ \tilde{\Phi} ^{2}}{4\widetilde{\Lambda }}\left[ dy^{3}+(\partial _{i}\ n)dx^{i}%
\right] ^{2}+\ \frac{(\partial _{4}\ \tilde{\Phi} )^{2}}{\ \widetilde{\Lambda }\
\Phi ^{2}}\left[ dy^{4}+(\partial _{i}\ \check{A})dx^{i}\right] ^{2}  \notag
\\
&&+\frac{\ ^{1}\tilde{\Phi}^{2}}{4\ ^{1}\widetilde{\Lambda }}\left[
dy^{5}+(\partial _{\tau }\ ^{1}n)du^{\tau }\right] ^{2}+\ \frac{(\partial
_{6}\ ^{1}\tilde{\Phi})^{2}}{\ ^{1}\widetilde{\Lambda }\ ^{1}\tilde{\Phi}^{2}%
}\left[ dy^{6}+(\partial _{\tau }\ ^{1}\check{A})du^{\tau }\right] ^{2}
\label{qellcs} \\
&&+\frac{\ ^{2}\tilde{\Phi}^{2}}{4\ ^{2}\widetilde{\Lambda }}\left[
dy^{7}+(\partial _{\tau _{1}}\ ^{2}n)du^{\tau _{1}}\right] ^{2}+\ \frac{%
(\partial _{8}\ ^{2}\tilde{\Phi})^{2}}{\ ^{2}\widetilde{\Lambda }\ ^{2}%
\tilde{\Phi}^{2}}\left[ dy^{8}+(\partial _{\tau _{1}}\ ^{2}\check{A}%
)du^{\tau _{1}}\right] ^{2},  \notag
\end{eqnarray}%
}where the generating functions $\Phi :=e^{\phi },\ ^{1}\Phi :=e^{\ ^{1}\phi
}$ and $\ ^{2}\Phi :=e^{\ ^{2}\phi }$ are re--parameterized, respectively, $%
\Phi \rightarrow \tilde{\Phi},$ $\ ^{1}\Phi \rightarrow \ ^{1}\tilde{\Phi},\
^{2}\Phi \rightarrow \ ^{2}\tilde{\Phi},$ following
\begin{eqnarray}
\frac{\partial _{4}[\Phi ^{2}]}{\ ^{v}\Lambda } &=&\frac{\partial _{4}[%
\tilde{\Phi}^{2}]}{\ \tilde{\Lambda}},\frac{\partial _{6}[\ ^{1}\Phi ^{2}]}{%
\ \ _{1}^{v}\Lambda }=\frac{\partial _{6}[\ ^{1}\tilde{\Phi}^{2}]}{\ \ ^{1}%
\tilde{\Lambda}},\frac{\partial _{4}[\ ^{2}\Phi ^{2}]}{\ \ _{2}^{v}\Lambda }=%
\frac{\partial _{4}[\ ^{2}\tilde{\Phi}^{2}]}{\ ^{2}\tilde{\Lambda}},  \notag
\\
\widetilde{\Lambda } &=&const,\ ^{1}\tilde{\Lambda}=const,\ ^{2}\tilde{%
\Lambda}=const.  \label{const2}
\end{eqnarray}%
The coefficients of metrics of type (\ref{qellcs}) satisfy the conditions
\begin{eqnarray}
s =0:\ && \tilde{\Phi}=\check{\Phi}(x^{i},y^{4},\mathring{\mu}),\partial
_{4}\partial _{k}\ \check{\Phi}=\partial _{k}\partial _{4}\ \check{\Phi};
\label{expconda} \\
&& \partial _{k}\ \check{\Phi}/\partial _{4}\ \check{\Phi} =\partial _{k}\
\check{A};\ _{1}n_{k}=\partial _{k}\ n(x^{i});  \notag \\
s = 1: && \ ^{1}\tilde{\Phi}=\ ^{1}\check{\Phi}(u^{\tau },y^{6},\mathring{\mu%
},\theta ),\partial _{6}\partial _{\tau }\ ^{1}\check{\Phi}=\partial _{\tau
}\partial _{6}\ ^{1}\check{\Phi};  \notag \\
&&\partial _{\alpha }\ ^{1}\check{\Phi}/\partial _{6}\ ^{1}\check{\Phi}%
=\partial _{\alpha }\ ^{1}\check{A};\ _{1}^{1}n_{\tau }=\partial _{\tau }\
^{1}n(u^{\beta },\mathring{\mu},\theta );  \notag \\
s=2: && \ ^{2}\tilde{\Phi}=\ ^{2}\check{\Phi}(u^{\tau _{1}},y^{8},\mathring{%
\mu},\theta ),\partial _{8}\partial _{\tau _{1}}\ ^{2}\check{\Phi}=\partial
_{\tau _{1}}\partial _{8}\ ^{2}\check{\Phi};  \notag \\
&&\partial _{\alpha _{1}}\ ^{2}\check{\Phi}/\partial _{8}\ ^{2}\check{\Phi}%
=\partial _{\alpha _{1}}\ ^{2}\check{A};\ _{1}^{2}n_{\tau _{1}}=\partial
_{\tau _{1}}\ ^{2}n(u^{\beta _{1}},\mathring{\mu},\theta ).  \notag
\end{eqnarray}%
The values $n(x^{i},\mathring{\mu},),\ ^{1}n(u^{\beta },\mathring{\mu}%
,\theta )$ and $\ ^{2}n(u^{\beta _{1}},\mathring{\mu},\theta )$ are
integrating functions which together with $\ \check{A}(x^{i},y^{4},\mathring{%
\mu},),\ ^{1}\check{A}(u^{\tau },y^{6},\mathring{\mu},\theta )$ and $\ ^{2}%
\check{A}(u^{\tau _{1}},y^{8},\mathring{\mu},\theta )$ determine the
N--connection coefficients (\ref{ncon}).

If conditions of type (\ref{const2}) are not imposed, we can consider
instead of $\partial _{i}\ \check{A},\partial _{\tau }\ ^{1}\check{A}%
,\partial _{\tau }\ ^{1}\check{A},$ respectively, the values
\begin{eqnarray*}
w_{i} &=&\partial _{i}\tilde{\phi}/\partial _{4}\widetilde{\phi }=\partial
_{i}\tilde{\Phi}/\partial _{4}\tilde{\Phi},\ ^{1}w_{\tau }=\partial _{\tau
}\ ^{1}\phi /\partial _{6}\ ^{1}\phi =\partial _{\tau}\ ^{1}\tilde{\Phi}%
/\partial _{4}\ ^{1}\tilde{\Phi}, \\
\ ^{2}w_{\tau _{1}} &=&\partial _{\tau _{1}}\ ^{2}\phi /\partial _{8}\
^{2}\phi =\partial _{\tau _{1}}\ ^{2}\tilde{\Phi}/\partial _{8}\ ^{2}\tilde{%
\Phi},
\end{eqnarray*}%
and construct exact solutions of (\ref{cdeinst}) which do not solve the
LC--conditions (\ref{lcconstr}). Such configurations are with
nonholonomically induced noncommutative torsion on $s=1,2$ shells.

The values (\ref{expconda}) contain dependence on integration parameters,
for instance, on a "commutative" one which can be related to the mass of
graviton $\mathring{\mu}$ and on "noncommutative" ones $\theta $ induced by
possible Schr\"{o}dinger type relations of type (\ref{ncom}). If the
LC--conditions (\ref{lcconstr}) are satisfied, the solutions of type (\ref%
{qellcs}) are for effective Einstein spaces with generic off--diagonal
metrics and polarizations of constants determined by massive gravity,
modifications of Lagrangians and certain noncommutative $\theta $%
--deformations. Such solutions may mimic violation of Lorentz symmetries and
HL--type anisotropy.

\section{On Renormalizable Modified Massive Gravity}

\label{sec4}

\subsection{Mimicking modified gravity and non--standard perfect fluid
coupling}

The solutions for theories for $~^{[1]}\mathcal{L}$ $\ $and $~^{[2]}\mathcal{%
L}$ are equivalent for classes of generating functions and sources
satisfying the conditions (\ref{const2}). Various models of modified gravity
with possible massive terms and bimetric/ biconnection structures can be
transformed into off--diagonal configurations of certain effective Einstein
spaces with shell cosmological constants $\widetilde{\Lambda },\ ^{1}\tilde{%
\Lambda},\ ^{2}\tilde{\Lambda}.$

The next step is to state the conditions when the solutions for a theory $%
~^{[2]}\mathcal{L}$ are encoded into a model $~^{[3]}\mathcal{L}.$ This
allows us to elaborate certain analogies of modified gravity to theories with
non--standard perfect fluid coupling \cite{odints1}. We work with generic
off--diagonal metrics (\ref{metr}) in 4--d (the constructions on (non)
commutative fibers can be performed via canonical lifts from the commutative
spacetime). With respect to coordinate frames and for a flat background
metric $\eta _{\alpha \beta },$ we write any\ $\mathbf{g}_{\alpha \beta }$ (%
\ref{metr}) as a generic off--diagonal metric $~^{\diamond }\mathbf{g}%
_{\alpha \beta }=\eta _{\alpha \beta }+h_{\alpha \beta }(x^{i},t),$ where $%
y^{4}=t$ is the timelike coordinate. Choosing the \textquotedblright
gauge\textquotedblright\ conditions $h_{tt}=h_{t\widehat{i}}=h_{\widehat{i}%
t}=0,$ for $\widehat{i},\widehat{j},=1,2,3$ for a \textquotedblright
double\textquotedblright\ splitting $(3+1)$ and $(2+2)$ on a manifold $%
\mathbf{V,}$ we express the corresponding Ricci tensor and scalar curvature
in the form {\small
\begin{eqnarray*}
~^{\diamond }R_{\widehat{i}\widehat{j}} &=&\frac{1}{2}(h_{\widehat{i}%
\widehat{j}}^{\ast \ast }+\partial _{\widehat{i}}\partial ^{\widehat{k}}h_{%
\widehat{j}\widehat{k}}+\partial _{\widehat{j}}\partial ^{\widehat{k}}h_{%
\widehat{i}\widehat{k}}-\partial _{\widehat{k}}\partial ^{\widehat{k}}h_{%
\widehat{i}\widehat{j}}),\ ^{\diamond }R_{44}=-\frac{1}{2}\delta ^{\widehat{i%
}\widehat{j}}h_{\widehat{i}\widehat{j}}^{\ast \ast }; \\
\ ^{\diamond }R &=&\delta ^{\widehat{i}\widehat{j}}(h_{\widehat{i}\widehat{j}%
}^{\ast \ast }-\partial _{\widehat{k}}\partial ^{\widehat{k}}h_{\widehat{i}%
\widehat{j}})+\partial ^{\widehat{i}}\partial ^{\widehat{j}}h_{\widehat{i}%
\widehat{j}}.
\end{eqnarray*}%
} For a prescribed generating function $\widetilde{\phi }(x^{i},t),\partial
_{t}\widetilde{\phi }\neq 0,$ there are elaborated models with $\ ~^{[2]}%
\mathcal{L}=~^{s}\widetilde{R}+\widetilde{L}=\ ^{[3]}\mathcal{L}=~^{\diamond
}R+~^{\diamond }L,$ with $~^{\diamond }L$ taken for $\ ^{\diamond }\mathbf{g}%
_{\alpha \beta }$ and an effective non--standard coupling with a fluid
configuration. Einstein manifolds encoding fluid like configurations are
generated if $\widetilde{\phi }$ contains parameters $\check{\alpha},\check{%
\beta},\rho ,\varpi $ introduced into equations%
\begin{eqnarray}
\frac{\partial _{t}\widetilde{\phi }\ \partial _{t}h_{4}}{2h_{3}h_{4}} &=&-\
^{\diamond }L=\check{\alpha}\rho ^{2}\{[\check{\beta}(3\varpi -1)+\frac{%
\varpi -1}{2}]\delta ^{\widehat{i}\widehat{j}}h_{\widehat{i}\widehat{j}%
}^{\ast \ast }+  \notag \\
&&(\varpi +3\varpi \check{\beta}-\check{\beta})(\partial ^{\widehat{i}%
}\partial ^{\widehat{j}}h_{\widehat{i}\widehat{j}}-\partial _{\widehat{k}%
}\partial ^{\widehat{k}}h_{\widehat{i}\widehat{j}})\}^{2}.  \label{ngenf}
\end{eqnarray}%
We write constants $\check{\alpha}$ and $\check{\beta}$ instead of $\alpha $
and $\beta $ in \cite{odints1}  to avoid possible ambiguities
related to coefficients $\alpha $ and $\beta $ from (\ref{genfc}). In above
formulas, we wrote $\ ^{\diamond }\Upsilon =-~^{\diamond }L$ in order to
emphasize that such a source is determined for a metric $~^{\diamond }%
\mathbf{g}_{\alpha \beta }$ and Lagrangian $~^{\diamond }L.$ Using this
expression and formulas (\ref{qellcs}) redefined following (\ref{const2}),
we find
 $h_{3}[\tilde{\Phi}]=e^{2\widetilde{\phi }}/4\widetilde{\Lambda }$ and $h_{4}[%
\tilde{\Phi}]=(\partial _{4}\widetilde{\phi })^{2}/\widetilde{\Lambda }$,
 which for $\tilde{\Phi}=e^{\widetilde{\phi }},$ $\ \widetilde{\Lambda }e^{2%
\widetilde{\phi }}=e^{2~^{0}\widetilde{\phi }}\int dt|~^{\diamond }\Upsilon
|\partial _{t}\left( e^{2\phi }\right) $ and $w_{i}=\partial _{i}\phi
/\partial _{t}\phi $ can be used for a $~\ ^{[1]}\mathcal{L}$--theory.

In the limit $\check{\beta}\rightarrow (1-\varpi )/2(3\varpi -1)$, we can
express $~^{\diamond }\Upsilon =\check{\alpha}\frac{\rho ^{2}}{4}(\varpi
+1)^{2}[(\partial ^{\widehat{i}}\partial ^{\widehat{j}}h_{\widehat{i}%
\widehat{j}}-\partial _{\widehat{k}}\partial ^{\widehat{k}}h_{\widehat{i}%
\widehat{j}})]^{2}$ and generate a class of Einstein manifolds
\begin{eqnarray}
\mathbf{g} &=&\epsilon _{i}e^{\psi (x^{k})}dx^{i}\otimes dx^{i}+\epsilon
_{3}\int dt|4\ ^{\diamond }\Upsilon |^{-1}\partial _{t}(e^{2\widetilde{\phi }%
})\mathbf{e}^{3}\otimes \mathbf{e}^{3}  \notag \\
&&+\epsilon _{4}\left[ \partial _{t}(\sqrt{|\int dt|~^{\diamond }\Upsilon
|^{-1}\partial _{t}(e^{2\widetilde{\phi }})|})\right] ^{2}e^{-2\widetilde{%
\phi }}\mathbf{e}^{4}\otimes \mathbf{e}^{4},  \label{sol2} \\
\mathbf{e}^{3} &=&dy^{3}+(\partial _{i}\phi /\partial _{4}\phi )dx^{i},\
\mathbf{e}^{4}=dy^{4}+n_{i}dx^{i},  \notag
\end{eqnarray}%
where $\epsilon _{\alpha }=\pm 1$ depending on signature. Such manifolds are
with Killing symmetry on $\partial /\partial y^{3}$ and broken Lorentz
invariance because \ the\ source $~^{\diamond }\Upsilon $ does not contain
the derivative with respect to $\partial /\partial y^{3}.$

It is not surprising that off--diagonal solutons of type (\ref{sol2}) are
not Lorentz invariant. For small nonholonomic deformations we can model, for
instance, rotoid Schwarz\-schild - de Sitter configurations which are
diffeomorphysm invariant and with broken Lorentz symmetry \cite{vexsolbranes}%
. In the ultraviolet region with large momentum $\mathbf{k}$, the second
term for the equivalent theory $\ ^{[3]}\mathcal{L}$ gives the propagator $|%
\mathbf{k|}^{-4}.$ For such configurations, the longitudinal modes do not
propagate being allowed propagation of the transverse one with possible
polarization of constant and additional off--diagonal terms. Alternatively,
we can say that we elaborated a theory with non--standard coupling of
modified/ massive gravity with perfect fluid when the energy--momentum
tensor $T_{\widehat{i}\widehat{j}}=p\delta _{\widehat{i}\widehat{j}}=\varpi
\rho \delta _{\widehat{i}\widehat{j}}$ and $T_{44}=\rho . $ Treating $p,\rho
$ and $\varpi $ as standard fluid parameters and the equation of state in
the flat background we compute $~^{\diamond }L=-\alpha (T^{\alpha \beta
}~^{\diamond }R_{\alpha \beta }+\beta T_{\alpha }^{\alpha }~^{\diamond
}R_{\beta }^{\beta })^{2}.$ The effective Ricci tensor can be considered for
$\nabla ,\ \widehat{\mathbf{D}}$ or any other metric compatible connection
completely defined by the matric structure following certain well defined
geometric principles.

The off--diagonal solutions for theories with $\ ^{[1]}\mathcal{L}$ and/or $%
\ ^{[2]}\mathcal{L}$ derived for generating functions and effective sources
of type (\ref{sourc1}) encode a kind of spontaneous violation of symmetry
which is typical in quantum field theories and condensed matter models. For
corresponding nonholonomic configurations, we elaborate configurations
vacuum gravitational aether via analogous coupling with non--standard fluid
which breaks the Lorentz symmetry for an effective equivalent theory $\
^{[3]}\mathcal{L}$ and mimic massive gravity contributions. This allows us
to elaborate a (power--counting) renormalizable model of QG, or to consider
other quantization procedures like deformation quantization, A--brane
quantization, gauge like gravity quantum models etc \cite{vquant}.

\subsection{Effective renormalizable (non)commutative modified and massive
gravity}

In this section, we consider a flat background approximation in order to study
equivalence of modified gravity theories (for certain classes of
nonholonomic constraints) to certain models with non--standard fluid
coupling. Such constructions can be extended to curved (in general,
noncommutative tangent bundles) backgrounds and generic off--diagonal
gravitational--field interactions using solutions of type (\ref{qellcs}).
Using generalizations of the principle of relativity \cite{vfinsl} (for the
so--called Einstein--Finsler models, noncommutative gravity etc), we can
work in a local N--adapted Lorentz frame when, for instance, the effective
fluid does not flow. This allows us to preserve unitarity and formulate
certain generalize axioms as in GR working with general class of solutions
for theories $\ ^{[1]}\mathcal{L}$ and/or $\ ^{[2]}\mathcal{L}.$ Choosing
corresponding classes of generating functions with necessary type parametric
dependence via corresponding nonholonomic transforms we mimic some models $\
^{[3]}\mathcal{L}$ with anisotropic coupling. The solutions of type (\ref%
{qellcs}) and (\ref{sol2}) in the diagonal spherical symmetry limit (here we
include the condition $T^{\alpha \beta }=0)$ contain the Schwarzschild and
Kerr black hole/ellipsoid metrics with various tangent bundle extensions
etc, see details in \cite{vexsolbranes}. Such configurations with $\Upsilon
=\ ^{\diamond }\Upsilon =-~^{\diamond }L$ (\ref{ngenf}) result in $z=2$ Ho%
\v{r}ava--Lifshitz theories with canonical (non) commutative extensions on
tangent Lorentz bundles.

The Lagrangian densities (\ref{act}) (for simplicity, we consider here 4--d
theories) include $z=3$ theories which allows us to generate ultra--violet
power counting renormalizable $3+1$ and/or $2+2$ quantum models. We can take
more general sources and generating functions instead of $~^{\diamond }L$
and write {\small
\begin{equation}
-\ _{\hat{n}}^{\diamond }L=\hat{\alpha}\{(T^{\mu \nu }\ ^{\diamond }\nabla
_{\mu }\ ^{\diamond }\nabla _{\nu }+\hat{\gamma}T_{\alpha }^{\alpha }\
^{\diamond }\nabla ^{\beta }\ ^{\diamond }\nabla _{\beta })^{\hat{n}%
}(T^{\alpha \beta }\ ^{\diamond }R_{\alpha \beta }+\hat{\beta}T_{\alpha
}^{\alpha }\ ^{\diamond }R_{\beta }^{\beta })\}^{2},  \label{auxa}
\end{equation}%
} where $\hat{n}$ and $\hat{\gamma}$ are constants and $\ ^{\diamond }\nabla
_{\mu }$ is obtained via nonholonomic constraints of any $\ ^{\diamond }%
\mathbf{D}_{\mu }$ which is metric compatible and completely determined by a
metric structure. Such configurations are determined by modified generating
functions as following. We use $\Upsilon =~-~_{\hat{n}}^{\diamond }L$ in
formulas (\ref{ngenf}) which results in a different class of generating
functions $~^{\hat{n}}\widetilde{\phi }$ from $\frac{\partial _{4}(~^{\hat{n}%
}\widetilde{\phi })\partial _{4}h_{3}}{2h_{3}h_{4}}=-~_{\hat{n}}^{\diamond
}L.$ A new class of off--diagonal solutions (\ref{sol2}) can be generated
for data $\widetilde{\phi }\rightarrow ~^{\hat{n}}\widetilde{\phi },\phi
\rightarrow ~^{\hat{n}}\phi $ and $~^{\diamond }\Upsilon \rightarrow -~_{%
\hat{n}}^{\diamond }L.$ Twisted (non) commutative extensions to higher
"velocity/ momentum" type fibers can be performed in a standard manner
as we considered in the previous section.  The formalism allows us to work
with models (\ref{auxa}) when $\hat{n}$ are non--integer, for instance, $\hat{n%
}=1/2,3/2$ etc (we do not study such theories in our works). We can apply
the analysis provided in \cite{odints1,vquant} which states that the
analogous $\ ^{[3]}\mathcal{L}$ and $\ ^{[4]}\mathcal{L}$ models derived for
(\ref{auxa}) are renormalizable if $\hat{n}=1$ and super--renormalizable for
$\hat{n}=2.$ Such properties can be preserved for theories generalized on
tangent Lorentz bundles.

Both on base spacetime manifold and on total space of corresponding Lorentz
bundles, the values induced by nontrivial (effective) sources and a
Lagrangian density $~_{\hat{n}}^{\diamond }L$ contain higher derivative
terms. Such terms break the Lorentz symmetry for high energies, i.e. in UV
region (which follows explicitly from noncommutative configurations). In the
IR limits, we can consider nonholonomic constraints when effective Einstein
configurations can be generated. Working with off--diagonal nonlinear
systems of PDE and their solutions, we can obtain the (effective) GR not
only as limits from certain modifications with non-standard coupling of type
(\ref{ngenf}) and/or (\ref{auxa}). We can model possible \textquotedblright
branches" of complexity, anisotropies, effective massive graviton
contributions, inhomogeneities and Lorentz violations depending on
parameters and generating functions. For well defined conditions, we can
select families of solutions which are (super) renormalizable because of
off--diagonal nonlinear interactions of gravitational and effective matter
fields. The models are, in general, with noncommutative quantum corrections.

\section{Concluding Remarks}

\label{sec5} Our results mean that using generic off--diagonal solutions for analogous (effective) gravity theories we can model various effects from modified gravity, massive gravity and various commutative and extra dimension generalizations on tangent bundles. The  main conclusion is that we can keep  an "orthodox" physical paradigm when the classical and quantum gravity theories are maximally closed to Einstein gravity. Our opinion is that the bulk of experimental data for modern gravity and cosmology can be explained/ predicted using generic off--diagonal solutions in general relativity (GR) and certain quantized locally anisotropic  versions on tangent Lorentz bundles. Such models can be encoded into nonholonomic Einstein manifolds and effective models of interactions with broken Lorentz symmetry and covariant renormalization.

In this paper, we do not prove explicitly the renormalizability of off--diagonal gravitational configurations but show that they can be associated/ related to certain "renormalizable" and ghost free quantum gravity models studied by other authors. The second main conclusion is that generic self--accelerating solutions in massive/ bigravity are anisotropic and can be alternatively modelled by modified gravity theories or, via nonholonomic constraints, as certain nonlinear off--diagonal interactions. Noncommutative models arise naturally on tangent Lorentz bundles if uncertainty quantum relations are included into consideration.

It would also be interesting to study off--diagonal deformations of Kerr black holes in modified massive gravity  and higher dimensions theories with (non) commutative velocity/ momentum type variables. Yet another interesting direction of research would be to study off--diagonal  ekpyrotic scenarios and possible equivalence of dark energy and dark matter models in modified,   massive and/or (effective) Einstein gravity.

\vskip5pt

\textbf{Acknowledgments:\ } The work is partially supported by the Program IDEI, PN-II-ID-PCE-2011-3-0256. Author is grateful to  S. Odintsov, P. Stavrinos, D. Vassilevich and M. Vi\c{s}inescu for important discussions and support. Some  preliminary results were communicated at the parallel section at MG13, Stockholm, 2012 (chairperson G. Amelino--Camelia).

\end{document}